\documentstyle[epsf.tex]{article}
\input math_macros
\font\tenbf=cmbx10
\font\tenrm=cmr10
\font\tenit=cmti10
\font\elevenbf=cmbx10 scaled\magstep 1
\font\elevenrm=cmr10 scaled\magstep 1
 1

\renewenvironment{thebibliography}[1]
 { \elevenrm
   \begin{list}{\arabic{enumi}.}
    {\usecounter{enumi} \setlength{\parsep}{0pt}
     \setlength{\itemsep}{3pt} \settowidth{\labelwidth}{#1.}
     \sloppy
    }}{\end{list}}

\begin{document}
\newcommand{\ltw}{\mbox{$\tilde{\Lambda}$}}
\newcommand{\shalf}{\mbox{$\scriptstyle \frac{1}{2}$}}
\epsfysize=1.5in
\setlength{\unitlength}{1in}

\begin{center}{{\tenbf APPLYING OPTIMIZED PERTURBATION THEORY TO \\
               \vglue 3pt
               QCD AT LOW ENERGIES\\}

\vglue 1.0cm
{\tenrm A. C. Mattingly and P. M. Stevenson\\}
\baselineskip=13pt
{\tenit
{T.W. Bonner Laboratory, Physics Department, Rice University, \\}}
\baselineskip=12pt
{\tenit {Houston, TX 77251, USA}\\}
\vglue 0.8cm
{\tenrm ABSTRACT}}
\end{center}
\vglue 0.3cm
{\rightskip=3pc
 \leftskip=3pc
 \tenrm\baselineskip=12pt
 \noindent
We discuss the use of the optimization procedure
based on the Principle of Minimal Sensitivity
to the third-order calculation of {\mbox{${R_{e^+e^-}}$}}.
The effective coupling constant
remains finite allowing us to
apply the Poggio-Quinn-Weinberg smearing method down to energies below 1 GeV,
where we find good agreement between theory and experiment.  The couplant
freezes to a value of $\alpha_s/\pi = 0.26$ at zero energy which is in
remarkable concordance with values obtained phenomenologically.
\vglue 0.5cm}
\baselineskip=14pt
\elevenrm

The
$e^+e^-$ annihilation cross section is a fundamental test of
QCD.  Recently
QCD corrections to third-order (NNLO) have been calculated using the
$\overline{\mbox{MS}}$ scheme \cite{prl}.  Here we use the method of
Ref.~\cite{opt}
to ``optimize'' the scheme choice.  It turns out this allows us to take
the calculations down to low energies.
$ R_{e^+e^-} $ has the form
\begin{equation}
 R_{e^+e^-} \equiv \frac{\sigma(e^+e^- \rightarrow \mbox{hadrons})}
{\sigma(e^+e^- \rightarrow \mu^+\mu^-)}
= 3\sum_q e_q^2 T(v_q) [1 + g(v_q)
\underbrace{a(1 + r_1 a + r_2 a^2 + \ldots)}_{\cal R} ]
\end{equation}
where $a \equiv \alpha_s/\pi$, and where the known functions $T(v_q),g(v_q)$
\cite{pqw} (equal to unity in the massless limit) allow for mass dependence
near thresholds.

A direct comparison of theory and data is not possible because of the
non-perturbative effects associated with thresholds;
instead we used the method of
Poggio,Quinn and Weinberg \cite{pqw} who define a ``smeared'' $R$:
\begin{equation}
\overline{R}_{PQW}(Q;\Delta) = \frac{\Delta}{\pi} \int_0^{\infty} ds^{\prime}
\frac{R_{e^+e^-}(\sqrt{s^{\prime}})} {(s^{\prime} - Q^2)^2 + \Delta^2}.
\end{equation}
The sharp resonances $\omega,\phi$,$J/\psi,\psi^{\prime}$,
and $\psi(3770)$ were left out of the data compilations
so that their contribution to $\overline{R}_{PQW}$
could be put in analytically.
In handling the experimental data we used both raw data points and eye-ball
fits to the data (especially in regions with a lot of structure).
Fitting the data allowed us to put in estimates for the experimental error.

A comparison between theory and experiment is given in Fig.~1
for $\Delta=2.0$ GeV$^2$.
Near the $\psi$ resonances some structure remains which
can be smoothed out by increasing $\Delta$.
Note that at low energies all structure has been smoothed out, and theory and
experiment agree very well.
\begin{figure}[hbt]
\epsfxsize= 6.0in
   \epsffile[72 288 540 504]{fig3.psf}
\tenrm\baselineskip=12pt \noindent
{Fig.~1: PQW-smeared $R$ ratio for $\Delta=2$ GeV$^2. \;\;\;\;\;$
Fig.~2: Couplant and
${\cal R}$ as a function of $Q$.}
\end{figure}
\elevenrm
\baselineskip=14pt
%

To get theoretical predictions down to such low energies we used ``optimized
perturbation theory '' (OPT),
based on the Principle of Minimal Sensitivity (PMS).  Here we just
sketch the ideas, for more detail see \cite{opt} and specifically \cite{us}.

\begin{picture}(6,.0001)(0,0)
 \put(4.90,2.65){\makebox(0,0){\Large ${\cal R}$}}
 \end{picture}
\vskip -.3in
Renormalization Group invariance means that a physical quantity ${\cal
R}$ is independent of the renormalization scheme (RS);
that is, it does not depend upon the particular way
one defines the renormalized couplant $a$.
Symbolically we can express this by:
\beq
 0 = \frac{d{\cal R}}{d(RS)} =  \frac{\partial {\cal R}}{\partial (RS)}   +
\frac{da}{d(RS)} \frac{\partial {\cal R}}{\partial a} \label{rg}
\eeq
where the total derivative can be separated into two pieces giving the RS
dependence via the coefficients and via the couplant.
The RS can be parameterized by a set of variables,
{$\tau,c_2,c_3,\ldots$} where
$\tau \equiv \ln(\mu/{\ltw})$ and $c_2,c_3,\ldots$ are the non-invariant
$\beta$ function
coefficients
\mbox{($
\beta(a) \equiv \mu \frac{da}{d\mu} =
-ba^2(1+ca + c_2a^2+\ldots)
$).}
Perturbative coefficients in the expansion of a physical
quantity can depend on the RS only through these parameters \cite{opt}.
The third-order approximant is of the form
\mbox{$ {\cal R}^{(3)} = a(1 + r_1 a + r_2 a^2)$} with a similar truncation of
the $\beta$ function.
The couplant then depends on just two RS parameters;
$a^{(3)} = a(\tau;c_2) $.
Eq.~(\ref{rg}) dictates how the coefficients $r_1,r_2$ must depend on
$\tau,c_2$.  One can express this by saying that the combinations
\begin{equation}
\rho_1 = \tau - r_1, \;\;\;\;\;\;\;\; \rho_2 = r_2 + c_2 - (r_1 + \shalf \;
c)^2
\label{inv}
\end{equation}
are RS invariants \cite{opt}.  Their values can be obtained from the
$\overline{MS}$ ($\mu=Q$) calculations of $r_1,r_2,c,c_2$.  [ $\rho_1$ depends
logarithmically on $Q$, while $\rho_2$ is a pure number, depending on $N_f$.]

While the exact ${\cal R}$ is RS-independent, the approximant ${\cal R}^{(3)}$
is not: it depends on $\tau$ and $c_2$ in a definite way.
The PMS criterion says that if an approximant depends on unphysical
parameters then those parameters
should be chosen so as to minimize the
sensitivity of the approximant to small variations in them \cite{opt}.
For ${\cal R}^{(3)}$ this implies
\begin{equation} \left. \frac{d {\cal R}^{(3)}}{d \tau}
\right|_{\bar{\tau},\bar{c_2}}  = 0,  \;\;\;\;\;\;\;\;\;\;\;\;
\left.  \frac{d {\cal R}^{(3)}}{d c_2} \right|_{\bar{\tau},\bar{c_2}}  = 0,
\end{equation}
providing two coupled equations determining
$\bar{\tau}$ and $\bar{c_2}$.  The integrated $\beta$ equation then determines
the optimized couplant, $\bar{a}$, given a $\Lambda$ value.  From
Eq.~(\ref{inv}) we can find the optimized coefficients
$\overline{r_1},\overline{r_2}$, and hence obtain the optimized ${\cal R}$.
Following this procedure gave the results shown in Fig.~2 for $\bar{a}$ and
${\cal R}$, as a function of $Q$.
\elevenrm
\baselineskip=14pt

The key result is that the optimized couplant does {\em{not}} go
through a pole at some finite $Q$ of order $\Lambda$.  Instead it ``freezes''
below 300 MeV to a nearly constant value.  This fixed-point behavior is due to
the large, negative $\overline{c_2}$ ($\approx -22$), which produces a
non-trivial zero in the OPT $\beta$ function.  The OPT analysis simplifies
considerably at a fixed point \cite{kubo}, yielding a simple formula for the
infrared
($Q\rightarrow 0$) limit of the optimized couplant:
\begin{equation}
{\scriptstyle \frac{7}{4}} +
c \bar{a}^* + 3 \rho_2 \bar{a}^{*2} = 0.
\end{equation}
For $N_f=2$ this gives
$a^{\ast} = 0.26$.

A ``freezing'' of $\alpha_s$ at low energies has long
been a popular assumption in
phenomenological models of low-energy hadronic physics \cite{pheno}.  This
phenomenology has been rather successful and extracts values for the low-energy
$\alpha_s/\pi$ that are remarkably close to ours.  Recently, an energy-loss
analysis of $b$-quark fragmentation has yielded a similar result \cite{vk}.  In
particular, comparison with the data yields 0.2 GeV as the integral of
$\alpha_s(k^2)/\pi$ over $k$ from 0 to 1 GeV \cite{vk}.
This happens to agree precisely with the
area under our curve in Fig.~2!

We stress that our value $a^{\ast}=0.26$ is determined purely by the
pertubative QCD calculations; it involves {\em{no}} experimental input
(not even the value of $\Lambda$).  This, we believe, is the first
purely {\em{theoretical}} evidence for the ``freezing'' of $\alpha_s$.
\vglue 0.5cm
{\elevenbf \noindent  Acknowledgements \hfil}
\vglue 0.3cm
We thank I. Duck, N. Isgur, and V. Khoze for helpful comments.
This work was supported in part by the U.S. Department of Energy under
Grant No. DE-FG05-92ER40717.
\vglue 0.5cm
{\elevenbf \noindent References \hfil}
\vglue 0.3cm

\end{document}